\documentclass[prd,preprintnumbers,amsmath,amssymb]{revtex4}
\usepackage{epsfig}
\begin{document}

\markboth{}
{\textit{D.A. Fagundes and M.J. Menon}}

\title{ Applicability of a Representation for the Martin's Real-Part Formula in
Model-Independent Analyses}

\author{D.A. Fagundes and M.J. Menon}


\affiliation{Universidade Estadual de Campinas - UNICAMP\\
Instituto de F\'{\i}sica Gleb Wataghin \\
31083-859 Campinas, SP, Brazil \\
fagundes@ifi.unicamp.br, menon@ifi.unicamp.br}

\begin{abstract}
Using a novel representation for the Martin's real-part formula without the
full scaling property, an almost model-independent description of the 
proton-proton differential cross section data at high energies 
(19.4 GeV - 62.5 GeV) is obtained.
In the impact parameter and eikonal frameworks,
the extracted inelastic overlap function presents a peripheral effect
(tail) above 2 fm and the extracted opacity function is characterized by a zero
(change of sign) in the momentum transfer space, confirming results
from previous model-independent analyses. Analytical parametrization
for these empirical results are introduced and discussed.
The importance of investigations on the
inverse problems in high-energy elastic hadron scattering is stressed
and the relevance of the proposed representation is commented.
A short critical review on the use of Martin's formula is also presented.
\keywords{Hadron-induced high- and super-high-energy interactions (energy $>$ 10 GeV);  elastic scattering.}
\end{abstract}

\maketitle

\vspace{0.3cm}

{\small
Keywords: 
Hadron-induced high- and super-high-energy interactions (energy $>$ 10 GeV); elastic scattering.}

{\small
PACS numbers: 
13.85.-t, 13.85.Dz.}

\vspace{0.5cm}

\centerline{\textit{Accepted for publication in International Journal of Modern Physics A, 2011}}

\vspace{0.5cm}

\textbf{Contents}

\vspace{0.2cm}

I. Introduction

\vspace{0.1cm}

II. Martin's Formula

\ \ \ II.A Original Derivation

\ \ \ II.B Geometrical Scaling, Developments and Critical Comments

\ \ \ II.C A Representation for Almost Model-Independent Analyses

\vspace{0.1cm}

III. Fit and Results

\ \ \ III.A Experimental Data

\ \ \ III.B Fit Procedures

\ \ \ III.C Results and Comments

\vspace{0.1cm}

IV. Extracted Overlap and Eikonal Functions

\ \ \ IV.A Impact Parameter and Eikonal Representations

\ \ \ IV.B Inelastic Overlap Function

\ \ \ IV.C Opacity Function in the Momentum Transfer Space

\vspace{0.1cm}

V. Conclusions and Final Remarks

\vspace{0.5cm}

\section{Introduction}	

The theoretical description of the high-energy \textit{elastic} hadron scattering
still constitutes a hard challenge for QCD: perturbative techniques
cannot be formally applied at this soft (large distances) region and 
nonperturbative approaches are not yet able to predict \textit{soft scattering states}
from first principles without model assumptions \cite{pred,land} some of them unjustified. 
On the other hand, in the phenomenological
context, a wide variety of models can describe the experimental data, but since they are characterized by
different physical pictures \cite{fiore} a comprehensive, well accepted and unified
formalism is still missing. This situation has to be contrasted with the great
expectations from the new experiments at the CERN LHC, in special the investigation
of the soft diffractive processes in proton-proton ($pp$) collisions by the 
TOTEM Collaboration \cite{totem1,totem2}.

At this stage, model-independent analyses, aimed to extract from the experimental data
empirical information on what is theoretically unknown, or on what cannot yet be evaluated,
may be an important strategy for realistic and necessary developments of the theory in the soft sector.
As first steps in that direction, we have investigated the elastic scattering 
based only on some general principles, high-energy theorems and model-independent fits to the
differential cross section data \cite{fms11,fms10,fm10,am,sma,acmm,cmm,cm}. Since the impact parameter and the unitarized eikonal representations
still constitutes a useful framework for model developments, the aims have been to extract
empirical information on the overlap and eikonal functions (impact parameter and momentum transfer
spaces).
However, since these analyses demand, as input, experimental information on the differential cross section
at intermediate and large values of the momentum transfer, they have been limited to proton-proton
scattering in the energy interval 19.4 - 62.5 GeV (as obtained from the CERN SPS, Fermilab and CERN ISR,
nearly forty years ago). Therefore, the new data from the LHC on pp collisions at 7 - 14 TeV,
covering the region up to large momentum transfer \cite{totem1,totem2} will certainly be
crucial for further developments.

One of the main problems with these model-independent analyses is the fact that the phase of the
scattering amplitude constitutes a physical observable only in the forward direction
(at zero momentum transfer), namely we do not have information on the contributions from the
real and imaginary parts of the amplitude beyond the forward direction. As a consequence, 
if we look for empirical analyses,
unbiased by any model/phenomenological ideas \cite{fiore} different kinds of contributions must be tested
and reliable information should not depend, or not strongly depend on the details of these contributions. 
With this strategy in mind, we have already discussed {\it two different empirical solutions} for
data reductions both statistically consistent \cite{fms11,am}. For example, the constrained results obtained in 
\cite{am} are in agreement with the standart picture, characterized by the dominance of the imaginary
part of the amplitude, except at the dip position, where the imaginary part presents a zero (change of sign)
and the dip is filled up by the real part. On the other hand, the recent
unconstrained results of \cite{fms11} indicate the dominance of the real part
at intermediate values of the momentum transfer,
just above the dip region.

Motivated by these different results, both consistent on statistical grounds, we consider here
a third possibility with different contributions from those obtained in the above mentioned previous analyses and developed by means
of a novel almost model-independent representation for what has been know as  Martin's real part 
formula \cite{martinf}.
This formula
connects real and imaginary parts of the elastic scattering
amplitude in terms of the transferred momentum in the collision; it has been derived
in the context of high-energy theorems and is embodied by
a scaling property.

In this work a representation for this formula, without the full scaling
property, is used in model-independent analysis of elastic $pp$ scattering at high energies.
By means
of an empirical parametrization for the imaginary part of the
amplitude, the use of the Martin's formula for the evaluation of the corresponding real part, and fits 
to $pp$ differential
cross section data (in the interval 19.4 - 62.5 GeV), the overlaps (elastic, inelastic and total) and
eikonal functions are extracted in an almost model-independent way.
The reproduction of
all the experimental data is quite good and the two main results are: the extracted 
inelastic overlap function (impact parameter space) is characterized by a peripheral effect (tail) above
2 fm and the extracted imaginary part of the
eikonal (opacity function) in the momentum transfer space
presents a zero (change of sign), in the region 6-7 GeV$^2$. Analytical parametrization for these
empirical results are introduced and discussed in certain detail.

In the context of our global investigation on the inverse problems in elastic hadron scattering \cite{fms11,am}
the focus here is only in the foundations, applicability
and some consequences of the proposed representation. With this aim we shall only outline references and discussions
related to results from previous analyses and phenomenological models. A global critical review,
taking into account results that are common to all the analyses (references \cite{fms11,am} and those presented
here), as well as detailed discussions on the implications in the phenomenological context will
be presented in a forthcoming work.

The manuscript is organized as follows. In Sect. II we review the original derivation of the Martin's
formula, present some critical comments on subsequent developments and introduce our representation. 
In Sect. III we describe the fit
procedures in some detail and display the results of the data reductions. In Sect. IV we treat the extracted quantities,
the inelastic overlap function and the eikonal in the momentum transfer space, making
reference to results
from previous analysis and  some phenomenological models. The conclusions and 
final remarks are the contents of Sect. V.

\section{Martin's formula}

General model-independent properties of the scattering amplitude can be obtained from
fundamentals principles of local quantum field theory and a set of 
rigorous theorems \cite{theo1,theo2,theo3,theo4};
in this context, unitarity, analyticity and crossing play a central role. Although rigorous,
some results are obtained under specific conditions, which define
and limit their range of validity and applicability. Martin's formula has been obtained
in this context and therefore it is important to stress the conditions involved.
In this Section we first treat the assumptions and main steps in the
original derivation \cite{martinf,martinmatthiae}, followed by a short critical review 
on the subsequent interpretations and applicability of the formula.
After that,
we introduce a representation for almost model-independent analysis, with a 
detailed discussion on the conditions involved.

\subsection{Original derivation}

For elastic processes, $m + m \rightarrow m + m$, the laboratory energy $E$ is a useful
crossing symmetry variable. Let $F$ be the scattering amplitude and $t$ the square of the
momentum transfer. In the high-energy limit ($E > > m$) the optical theorem reads \cite{bc}
\begin{eqnarray}
\sigma_{\mathrm{tot}} = 4 \pi \frac{\mathrm{Im}F(E, t=0)}{E},
\nonumber
\end{eqnarray}
and the Froissat-Martin bound states that \cite{fromar1,fromar2,fromar3}
\begin{eqnarray}
\sigma_{\mathrm{tot}} \leq C [\ln E]^2,
\nonumber
\end{eqnarray}
with $C$ a constant.

In what follows we shall use the symbol $\sim$ for an asymptotic equality to with a constant factor.
Auberson, Kinoshita and Martin \cite{akm} (AKM) have shown that if the above bound
is saturated ($\sigma_{\mathrm{tot}} \sim [\ln E]^2$), then for
\begin{eqnarray}
|t| < \frac{k}{[\ln E]^2},
\quad
k\ \mathrm{arbitrary\ constant},
\end{eqnarray}
 an scaling property holds
\begin{eqnarray}
\dfrac{F(E,t)}{F(E,0)} \rightarrow f(\tau),
\qquad 
\tau= t[\ln E]^2
\end{eqnarray}
and therefore, in that case, from the optical theorem,
\begin{eqnarray}
F(E,t) \sim E [\ln E]^2 f(t[\ln E]^2).
\nonumber
\end{eqnarray}

Real and imaginary parts of the amplitude can be correlated through high-energy theorems based on
analyticity and the crossing relations established by Bros, Epstein and Glaser in the context
of quantum field theory \cite{beg1,beg2}. The main ingredient is the condition of polynomial boundedness and the
Phragm\'en-Lindel\"off theorem which leads to the asymptotic uniqueness \cite{eden}.
By assuming that the odd-under-crossing amplitude can be neglected
at high energies, exact crossing for the even ($+$) amplitude demands that the above real amplitude 
has structure \cite{martinmatthiae,eden}

\begin{eqnarray}
F_+ \sim \mathrm{i}E[\ln E-\mathrm{i}\dfrac{\pi}{2}]^{2}f(t[\ln E-\mathrm{i}\dfrac{\pi}{2}]^2).
\nonumber
\end{eqnarray}
Since $[\ln E-\mathrm{i}\dfrac{\pi}{2}]^2 \approx [\ln E]^2 - \mathrm{i} \pi \ln E$, the function $f$
can be expanded and in first order we get
\begin{eqnarray}
f(\tau - \dfrac{\mathrm{i}\pi \tau}{\ln E}) \approx f(\tau) - \dfrac{\mathrm{i}\pi \tau}{\ln E} \dfrac{df}{d\tau},
\nonumber
\end{eqnarray}
so that in the high-energy limit the Martin's result is obtained
\begin{eqnarray}
\mathrm{Im}F_+(E,t) &\sim & E[\ln E]^{2}f(\tau), \\
\mathrm{Re}F_+(E,t) &\sim & \pi E\ln E \dfrac{d}{d\tau}(\tau f(\tau)).
\end{eqnarray} 
Hence, for a given $f(\tau)$ (imaginary part), the real part can be evaluated. We stress the 
four main conditions in this derivation:
(1) asymptotic limit (Froissart-Martin bound reached); (2) even amplitude; 
(3) scaling property; (4) $|t| < k/[\ln E]^2$.

\subsection{Geometrical Scaling, Developments and Critical Comments}

\begin{quote}
``Sometimes these theorems called `asymptotic' are not expected to apply to present
available energies, but in fact some of them do apply already now. Then they are no longer theorems 
but they give useful trends."
\end{quote}
\centerline{A. Martin, G. Matthiae \cite{martinmatthiae}}

\vspace{0.2cm}

In what follows we shall consider the center-of-mass system, with focus in this section
on $pp$ and $\bar{p}p$
elastic scattering. Let $A(s, q)$ be the complex amplitude,  where $s$ is the square of the energy
and $q^2 \equiv - t$. The physical quantities of interest here, with the corresponding normalizations,
are the differential cross section
\begin{eqnarray}
\frac{d\sigma}{dq^{2}} = \pi |A(s,q)|^{2},
\end{eqnarray} 
the total cross section
\begin{eqnarray}
\sigma_{\mathrm{tot}} = 4\pi \mathrm{Im} A(s,0),
\end{eqnarray} 
and the $\rho$ parameter,
\begin{eqnarray}
\rho(s)=\dfrac{\mathrm{Re}A(s,0)}{\mathrm{Im} A(s,0)}.
\end{eqnarray} 

Martin's result (3-4) was published in 1973, a few months after the introduction of the
Geometrical Scaling (GS) by Dias de Deus \cite{dd}. This phenomenological approach,
further developed by Dias de Deus, Buras and Kroll \cite{gs1,gs2,gs3} was based on the 
\textit{empirical
evidence} of a scaling in elastic $pp$ scattering at the ISR energy region
($\approx 20 - 60$ GeV) in terms of the variables
\begin{eqnarray}
\frac{1}{\sigma_{\mathrm{tot}}^2}\,\frac{d\sigma}{dq^{2}} = \Phi(q^2 \,\sigma_{\mathrm{tot}}).
\end{eqnarray} 
The geometrical aspect concerns the fact that for $\sigma_{\mathrm{tot}}(s) \propto R^2(s)$,
with $R(s)$ an effective interaction radius, the increase of the
total cross section and all the energy dependence involved 
in $d\sigma/dq^2$ is only due to an expansion (geometrical) effect of the hadron.

Now, since from first order derivative dispersion relations \cite{ddr1,ddr2,ddr3} and for an even amplitude,
\begin{eqnarray}
\rho(s) \approx \frac{\pi}{2 \sigma_{\mathrm{tot}}(s)} \frac{d}{d\ln s}
\sigma_{\mathrm{tot}}(s),
\end{eqnarray} 
the Martin results, Eqs. (3) and (4), along with normalizations (5), (6) and (7), can be put
in the form

\begin{eqnarray}
\frac{d\sigma}{dq^2} = {\frac{\sigma_{\mathrm{tot}}^2}{16\pi}
\left\{ \left[\rho \frac{d}{d\tau} [\tau f(\tau)]\right]^2
+ \left[f(\tau)\right]^2\right\}},
\end{eqnarray}
which also characterizes the GS model \cite{gs1,gs2,gs3}, Eq. (8), for $\tau = q^2 \,\sigma_{\mathrm{tot}}$
(not necessarily asymptotic energies).

Therefore, the result expressed by Eq. (10) (hereafter referred to as Martin's formula,
as in the literature), has two independent foundations:

\begin{itemize}
\item[(1)] the AKM scaling, Eq. (2), a rigorous formal result, deduced under strictly 
conditions of asymptotic energies (Froissart-Martin bound saturated), valid in a limited
interval of momentum transfer, Eq. (1) and for even amplitude;
\item[(2)]  the GS, an empirical result verified at the ISR energy region (certainly not
asymptotic), without restriction in momentum transfer or crossing properties of
the amplitude.
\end{itemize}

That seems a peculiar and/or accidental  fact, putting in evidence some intrinsic
difficulties associated with the asymptotic concept and the elastic processes
in hadronic interactions, in general, as commented by Martin and Matthiae \cite{martinmatthiae}.
Let us shortly review some further developments concerning the energy and momentum transfer
variables associated with the Martin's formula.

After the introduction of these scaling
in the seventies,
the new energy domain reached with subsequents experiments on $\bar{p}p$
scattering at the CERN S$\bar{p}p$S collider (546 GeV) and Fermilab Tevatron (1.8 TeV),
have shed light on some aspects involved, as follows.
The GS predicts a constant value for the ratio between the elastic and total cross 
section, a result in agreement with the experimental data at the ISR energy region.
However, at the Collider and Tevatron energies it has been verified that
this ratio increases \cite{bozzo}, leading to the breakdown of the GS. On the other hand,
in the phenomenological context, Henzi and Valin have shown that from the ISR
to Collider, the energy evolution of the dip-bump structure of the differential
cross section demands not only a geometrical expansion but also blackening and
edging effects \cite{hv1,hv2}. With their BEL approach (Blacker, Edgier, Larger),
based on dispersive diffraction theory, corrections have been introduced in the
Martin's formula, leading to extensions at higher energies \cite{hv-high}.

Another aspect concerns the interval in momentum transfer in which Eq. (10) is expected
or supposed to be valid and that plays a central role in our analysis. In this
respect and in the context
of the AKM scaling, Kundr\'at and Lokaj\'{\i}\v{c}ek have developed a detailed numerical 
analysis \cite{kl1,kl2,kl3} showing that the function $f(\tau)$ in (10) is real only in a limited interval of 
the scaling variable $\tau$, corresponding to the region in momentum transfer
0 $\leq q^2 \leq$ 0.15 GeV$^2$. Therefore, the use of the Martin's
formula outside this region has no physical meaning. This conclusion, however, has been 
criticized by Kawasaki, Maehara and Yonezawa \cite{kmy-c} and according to these authors,
Martin's formula can be directly deduced through derivative dispersion relations
in the case that GS holds, which is verified, at least, at the ISR energy region.
Therefore, in this energy domain there would be no limit for the applicability
of the Martin's formula in terms of the momentum transfer.

Despite the above discussion, a crucial result for our purposes has been introduced
by S.M. Roy \cite{roy} in the investigation of unitarity inequalities connecting 
Re $A(s, q)$ and
Im $A(s,q)$. The main ingredient, suggested by the author
and raised by A. Martin (see acknowledgments in \cite{roy}), concerns the possibility 
of a non-scaling only in the imaginary part of the amplitude. In fact, if Re $A$ is small
compared with Im $A$ it seems reasonable to consider the scaling property only
in the evaluation of the real part. The proposal by Roy has been to consider Eq. (10)
but now with the substitution
\begin{eqnarray}
f(\tau) \Rightarrow g(s,q) = \frac{\mathrm{Im}\, A(s,q)}{\mathrm{Im}\, A(s,0)}.
\qquad
\end{eqnarray}
that is, the scaling is supposed to hold only in the small contribution of the real part.

Although under limited formal condition of validity, this proposal can be
assumed as a representation for the Martin's formula applicable, in principle,
to all values of $s$ and $q^2$. In fact, despite its
``hybrid" character the efficacy of this representation
in the phenomenological context has been demonstrated, subsequently, by several authors, for
both $pp$ and $\bar{p}p$ scattering and different regions of the momentum transfer
and energy with experimental data 
available \cite{bcmp,malecki,menon1,cy1,cy2,menon2,menon3,kmy1,kmy2,ddp}.

\subsection{A representation for almost model-independent analyses}

As commented in our introduction, motivated by two different model-independent
results for the contributions from the real and imaginary parts of the amplitude \cite{fms11,am}
we consider here a third
possibility, based on a representation for the Martin's formula. The point
is the Roy's proposal, Eq. (11), with a model-independent parametrization
for the imaginary part of the amplitude, in the usual form of a sum of Gaussian
in $q$ (or exponential in $q^2$). Specifically, in Eq. (11) we consider
\begin{eqnarray}
g(s,q) 
\equiv
\sum_{i=1}^{n} a_i e^{- b_i q^{2}},
\end{eqnarray}
where $a_i$ and $b_i$, $i = 1, 2,....,n$ are real free parameters.
With this assumption and through Eq. (10), we obtain the following representation for the full
complex amplitude
\begin{eqnarray}
A(s,q) &=& \left\{ \left[\frac{\rho \ \sigma_{\mathrm{tot}}}{4 \pi \sum_{i=1}^{n}a_i} \right]
\ \frac{d}{dq^2}
\left[q^2 {\sum_{i=1}^{n} a_i e^{- b_i q^{2}}} \right]\right\} \nonumber \\ 
&+& \mathrm{i} \left\{
\left[\frac{\sigma_{\mathrm{tot}}}{4 \pi \sum_{i=1}^{n}a_i}\right] 
{\sum_{i=1}^{n} a_i e^{- b_i q^{2}}}\right\}. 
\end{eqnarray}

Using as input the experimental values of $\sigma_{\mathrm{tot}}$ and $\rho$ in each energy analyzed,
we shall consider fits to the differential cross section data through Eqs. (5) and (13). We note that
in this representation $s$ and $q^2$ are independent variables and the number of free
parameters is $2n - 1$, due to the constraint
\begin{eqnarray}
\sum_{i=1}^{n} a_i = 1.
\end{eqnarray}
However, before considering the above parametrization in our data analysis, some critical
comments on its applicability here are in order:

\begin{itemize}
 \item The analysis by Kundr\'at and Lokaj\'{\i}\v{c}ek \cite{kl1,kl2,kl3} concerns the scaling variable
$\tau = q^2\ln^2 s$ or $\tau = q^2 \sigma_{\mathrm{tot}}$. Since in the representation  (13) the variables
$s$ and $q^2$ are independent (Roy's proposal), we understand that the restriction to the narrow interval in the momentum transfer, referred to in Sect. II.B, does not apply.
\item As told before, we shall treat only $pp$ scattering in the energy interval 
19.4 GeV $\leq \sqrt{s} \leq$ 62.5 GeV, a region (ISR) where the GS is empirically verified \cite{as}.
Therefore, we see no restriction on its use in the evaluation of the real part of the amplitude.
\item On the other hand, just the above energy interval brings about some shortcomings related to
formal and practical aspects involved. In fact, at the ISR region, $pp$ and $\bar{p}p$ scattering
are distinct in what concerns the experimental data on $d\sigma/dq^2$, $\sigma_{\mathrm{tot}}$ and $\rho$.
Therefore the neglecting of the odd-crossing amplitude (connected with the asymptotic limit in the
Martin's derivation) cannot be justified. Now, even if the odd (-) contribution could
be taken into account, that would demand a simultaneous analysis of $pp$ and $\bar{p}p$ scattering,
since
\begin{eqnarray}
A_{\pm} = \frac{A_{\bar{p}p} \pm A_{pp}}{2}.
\end{eqnarray}
However, in that case model-independent fits cannot be statistically developed
due to the small interval in momentum transfer with data available on $\bar{p}p$
scattering \cite{cmm}.
\end{itemize}

Therefore, despite the advantages of parametrization (13), we
understand that it has a limited formal
justification in what concerns the evaluation of the real part of the
amplitude for $pp$ scattering in the energy interval of interest here.
That was the reason why we refer to an almost model-independent analysis.

On the other hand, we stress that we do not intend to extract any direct physical 
implication or interpretation
from the specific contribution of the real part of the amplitude. On the contrary, as commented
in the introductory section, our strategy is to look for global properties from
different parametrization and fit results that are not connected with or do not
strongly depend on the (unknown) contributions from Re$A$ and Im$A$ beyond the forward direction.
With that in mind and as we shall show in the following sections,
parametrization (13) brings new useful insights in the investigation of the
inverse problems in elastic hadron scattering.

\section{Fit and results}

Our purpose is to use parametrization (13) to fit
differential cross section data through Eq. (5), using as input the
experimental values on $\sigma_{\mathrm{tot}}$ and $\rho$ in each set analyzed.
In this section, we first refer to the data set in which our analysis is based
(Sect. III.A), then we discuss the fit procedures in certain detail (Sect. III.B) and
after that we present the fit results and some critical comments (Sect. III.C).

\subsection{Experimental data}

In the context of our empirical analyses \cite{fms11,fms10,fm10,am,sma,acmm,cmm,cm}, the evaluation of the uncertainties
in the extracted quantities plays a central role. Since the fits are based
on the differential cross sections, that demands sets with available data
covering the largest intervals in momentum transfer, mainly in what concerns
the region of high momentum transfer. As already discussed in some
detail \cite{fms11,am,cmm} this condition limits the analysis to six
sets of $pp$ elastic scattering data, at $\sqrt{s}$ = 19.4 GeV (from the Fermilab and CERN
SPS) and  $\sqrt{s}$ = 23.5, 30.7, 44.7, 52.8 and 62.5 GeV (from the CERN ISR).
In addition, we make use of the empirical result that at the ISR energy region (23.5 - 62.5 GeV)
the differential cross section data above $q^2 \sim$ 4 GeV$^2$ do not depend on 
the energy \cite{nagy,faissler,dl1,dl2}.
This fact allows the inclusion in each set (ISR) the data at 27.4 GeV (from Fermilab), covering the
large momentum transfer region: 5.5 $\leq$ q$^2$ $\leq$ 
14.2 GeV$^2$ (see \cite{am} for a quantitative discussion on this respect).

The data set have been collected from final published results
and a complete list of references, with comments, can be found
in \cite{fms11} for the data at 19.4 GeV and in \cite{am}
for the ISR data (23.5 - 62.5 GeV). The differential cross sections data
include the optical point,
\begin{eqnarray}
\left.\frac{d\sigma}{dq^2} \right|_{q^2=0} = 
\frac{\sigma_{\mathrm{tot}}^2 (1 + \rho^2)}{16\pi} 
\end{eqnarray}
and the data above the Coulomb-nuclear interference region, $q^2\ >\ $ 0.01 GeV$^2$.
All the available data have been included, without any kind of selection or
exclusion of data points.
As a test for goodness of fit we shall consider the $\chi^2$ per degrees of freedom (DOF)
and since this test is based on the assumption of a Gaussian error distribution,
only the statistical errors of the experimental data are taken into account (we do not include systematic errors).
We shall return to this point in our comments at the end of Sect. III.C.

\subsection{Fit procedures}

The data reductions have been performed using the CERN-Minuit code \cite{minuit} through successive 
runs of the MIGRAD minimizer and with the confidence level for the
uncertainties in the free parameters fixed at 70 \% . The nonlinearity of the fit
demands the start values of the free parameters and 
looking for a completely unbiased procedure the following methodology has been used.

We have begun the analysis with the
differential cross section set corresponding to the largest 
experimental information available, namely $\sqrt{s}$ = 52.8 GeV
(plus data at 27.4 GeV, as explained in Sect. III.A) and through the
following steps:

\begin{itemize}

\item[1.] First we have considered only the optical point and the data in
the narrow interval 0.01 $< q^2 \leq$ 0.1 GeV$^2$
(very forward direction). In the logarithmic scale, the differential cross
section data follows a well defined straight line, so that the
slope and the intercept can be directly evaluated from the plot
and used as start values for the fit with only one exponential term
in parametrization (13), that is, $n$ = 1.

\item[2.] Once obtained the final fit result,
we have enlarged the interval up to $q_{\mathrm{max}}^2$ = 0.2 GeV$^2$. Due to 
a sudden change of the slope around 0.13 GeV$^2$, we have considered two exponential terms
($n$ = 2) and used as start values for the parameters in both terms the final values of the previous fit. 
With this procedure the break of the slope is quite well described, as well as all the
data in the above interval.

\item[3.] Next we have considered $q_{\mathrm{max}}^2$ = 0.5 GeV$^2$ and this case demands
$n=3$. As start values for the third exponential terms we have tested the final
values of each of the two terms in the previous step and selected the best
result (reduced $\chi^2$ closest to 1).

\item[4.] The same procedure has been then applied for  $q_{\mathrm{max}}^2$ = 1.0, 2.0, 3.0, 4.0  GeV$^2$
and after that, $q_{\mathrm{max}}^2$ = 14.2 GeV$^2$.

\end{itemize}

Once obtained the best result at 52.8 GeV, the final values of the parameters have been 
used as feed back for the fits at each nearby energy set, namely from 52.8 GeV to
62.5 GeV, from 52.8 GeV to 44.7 GeV and then from 44.7 GeV to 30.7, 23.5 and 19.4 GeV,
successively.

As described above, we did not impose any kind of constraint in the fit procedure,
in special in what concerns the number of parameters or possible dependencies
of the parameters with the energy. The only goal has been to obtain the best fit
on statistical grounds.

\subsection{Results and Comments}

The final values of the free parameters and statistical information on the
fit results for each set analyzed are displayed in Table I.
With the Minuit code, the error-matrix provides the variances and covariances associated with
each free parameter. This information, together with the statistical errors of
$\sigma_{\mathrm{tot}}$ and $\rho$, are used in the evaluation of uncertainty regions in the
differential cross section and all the extracted quantities, through standard error propagation
procedures \cite{bevington}.
The results together with the corresponding experimental
data are shown in Fig. 1 in all the $q^2$ region and at the diffraction peak.
The contributions to the differential cross sections from the real and imaginary parts
of the amplitude,
\begin{eqnarray}
\frac{d\sigma^R}{dq^2} = \pi\ [\mathrm{Re}\, A]^2, 
\qquad
\frac{d\sigma^I}{dq^2} = \pi\ [\mathrm{Im}\, A]^2,
\end{eqnarray}
are shown in Fig. 2, where
we display only the uncertainty regions and the experimental data.

\begin{table}[!]
\caption{Fit results for each data set and statistical information: degrees of freedom
(DOF) and reduced chi square ($\chi^2/$DOF). For i=1,2,...,6, 
the parameters $a_i$ are dimensionless and $b_i$ in GeV$^{-2}$.}
{\begin{tabular}{@{}ccccccc@{}}
\toprule
$\sqrt{s}$ (GeV):  & 19.4  & 23.5   & 30.7   & 44.7   & 52.8  & 62.5 \\
\colrule
$a_1$        &  0.111 & 0.506  & 0.333  & 0.0791  & 0.151 & 0.179   \\
	           & $\pm$0.011 & $\pm$0.029 & $\pm$0.019  & $\pm$0.0097 & $\pm$0.017 & $\pm$0.022  \\
$a_2$        &  5.150 & 5.734  & 5.740 & 5.825 & 6.8323 & 6.7449   \\
                   & $\pm$0.014 & $\pm$0.032 & $\pm$ 0.023& $\pm$0.017 & $\pm$0.018 & $\pm$0.019   \\
$a_3$        &  -8.4565 &  -8.4602 & -8.4491 &  -8.444 &  -7.5697  & -7.6620  \\ 
	           & $\pm$0.016 & $\pm$0.033 & $\pm$0.024 & $\pm$0.017 & $\pm$0.018 & $\pm$0.019\\
$a_4$        & -0.985 &  -0.00665  & -0.0089 & -0.0099  & -0.00862 & -0.0064 \\
	           & $\pm$0.011 & $\pm$0.00085 & $\pm$0.0011  & $\pm$0.0012 & $\pm$0.00082& $\pm$0.0012 \\
$a_5$        &  -0.00165 & -0.000142  & -0.000271 & -0.00037 & -0.00023 & -0.000203   \\
                   & $\pm$0.00010 & $\pm$0.000069 & $\pm$ 0.000090& $\pm$0.00011 & $\pm$0.000075 & $\pm$0.000093   \\
$a_6$        &  5.187 &  3.219 &  3.384 &  3.564 &  1.598  & 1.742   \\ 
	           & $\pm$0.017 & $\pm$0.036& $\pm$0.027 & $\pm$0.018 & $\pm$0.019 & $\pm$0.020\\
$b_1$         &   14.7  &  7.82  &   9.56  &  31.3 & 16.4  & 14.1   \\ 
	           & $\pm$1.7 & $\pm$0.45& $\pm$ 0.60  & $\pm$5.8 & $\pm$1.8 & $\pm$1.7\\
$b_2$         &   3.2466  & 3.3610 & 3.3095 &  4.1790 & 3.9811   & 3.786   \\ 
	           & $\pm$ 0.0045 & $\pm$0.0061 & $\pm$0.0040  & $\pm$0.0085 & $\pm$0.0068 & $\pm$0.013\\
$b_3$         &   3.7449 & 3.4191 &  3.4232   &  4.4031 & 4.0619   & 3.852   \\ 
	           & $\pm$0.0045 & $\pm$0.0033 & $\pm$0.0083 & $\pm$0.0078& $\pm$0.0069 & $\pm$0.013\\
$b_4$         &   2.7230  & 1.004  & 1.150  &  1.242 & 1.156  & 1.079   \\ 
	           & $\pm$0.0095 & $\pm$0.055& $\pm$0.057  & $\pm$0.067 & $\pm$0.046 & $\pm$0.079\\
$b_5$         &   0.5993  & 0.373   & 0.433  &  0.469 & 0.426   & 0.418   \\ 
	           & $\pm$0.0084 & $\pm$0.046 & $\pm$0.036  & $\pm$0.035 & $\pm$0.036 & $\pm$0.046\\
$b_6$         &   4.277 & 3.610 &  3.8146 &  5.121 & 5.1138   & 4.756   \\ 
	           & $\pm$0.010 & $\pm$0.015 & $\pm$0.013 & $\pm$0.025 & $\pm$0.057 & $\pm$0.078\\
DOF           & 302  & 161   & 200   & 235   & 233   & 152\\
$\chi^2/$DOF    & 3.00 & 1.59  & 1.42  & 2.11  & 1.74  & 1.23\\
\botrule
\end{tabular}
\label{ta1}}
\end{table}

\begin{figure}[pb]
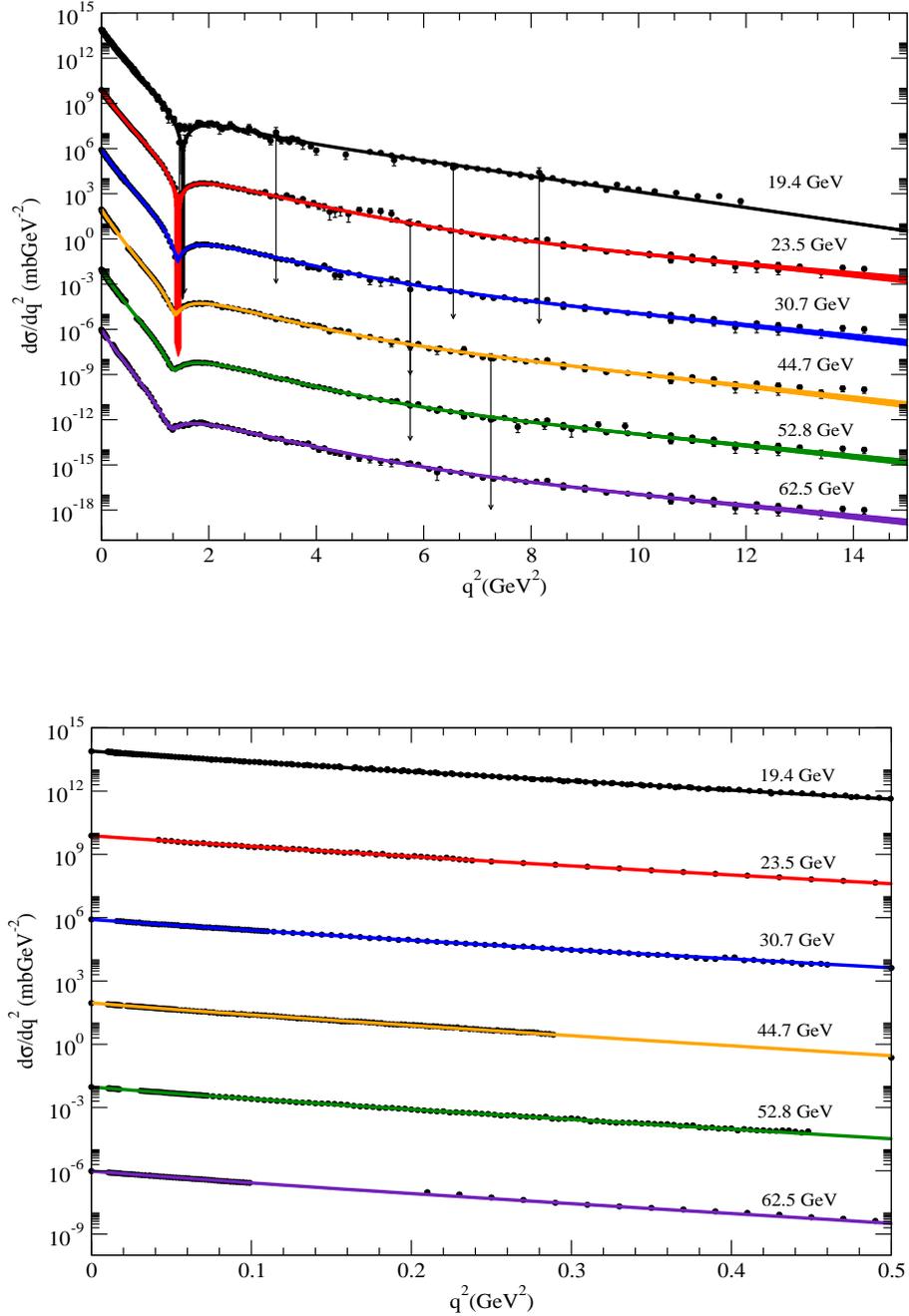

\centering
\vspace*{1.0cm}
\epsfig{file=f1a.eps,width=12cm,height=8cm}\vspace*{1.5cm}
\epsfig{file=f1b.eps,width=12cm,height=8cm}
\vspace*{8pt}
\caption{Fit results and uncertainty regions from error propagation:
all $q^2$ region (above) and diffraction peak (below). Curves and data
have been multiplied by factors of $10^{\pm 4}$.
\label{f1}}
\end{figure}

\begin{figure}[pb]
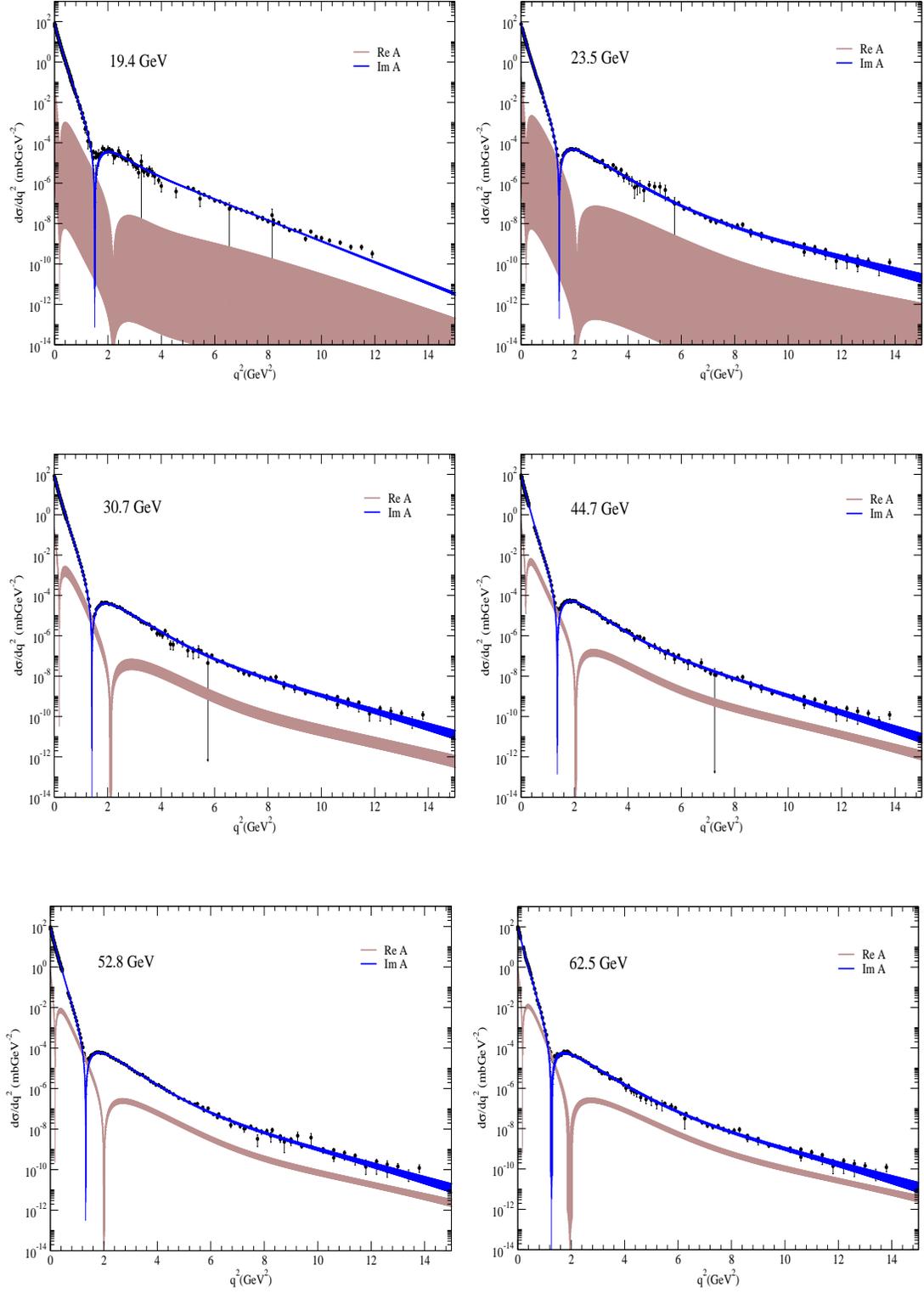

\centering
\vspace*{0.49cm}
\epsfig{file=f2a.eps,width=7cm,height=6cm}\hspace*{0.2cm}
\epsfig{file=f2b.eps,width=7cm,height=6cm}\\
\vspace*{1.05cm}
\epsfig{file=f2c.eps,width=7cm,height=6cm}\hspace*{0.2cm}
\epsfig{file=f2d.eps,width=7cm,height=6cm}\\
\vspace*{1.05cm}
\epsfig{file=f2e.eps,width=7cm,height=6cm}\hspace*{0.2cm}
\epsfig{file=f2f.eps,width=7cm,height=6cm}
\vspace*{4pt}
\caption{Experimental data and uncertainty regions in the contributions to the 
differential cross sections from the real (gray) and imaginary (black) parts of the amplitude.
\label{f2}}
\end{figure}

Before considering the extracted quantities some comments on the fit
results presented in this section are in order, as follows.

From Table I, although the values of each parameter in each one of the six sets analyzed are practically
of the same order of magnitude, a dependence on the energy cannot be inferred.
That is a consequence of our fit procedure, developed without any kind of
constraint in the free parameters, as well as the relatively small energy
interval investigated ($\approx$ 19 - 60 GeV). 
In fact, the same undefined energy dependence on the free parameters characterizes
the results in our previous analyses, with both the constrained \cite{am} and
unconstrained \cite{fms11} parametrization.
In this respect, the new data on the differential
cross section from the LHC (7 - 14 TeV) reaching the large momentum transfer region
may bring important information for further investigation
on possible energy dependencies.

Although Figure 1 shows that all the experimental data are quite well described, 
from Table I we  note that the values of  the $\chi^2/$DOF for
about 200 DOF correspond to extremely small confidence intervals.
That is a consequence of the fact that we have considered only statistical errors
and not the systematic ones. The inclusion of the systematic errors
(for example in quadrature with the statistical ones) lead to quite good
confidence intervals, but we do not think this procedure has  statistical
meaning, as commented at the end of in Sect. III.A.

From the plots in Figure 2, the contributions to the differential cross section from Re $A$ and
Im $A$ are in agreement with the standard or generally expected behavior: dominance
of the contribution from Im $A$ except at the dip region, which is filled up by the contribution
from Re $A$. In all sets Im $A$ present a zero (change of sign) at the dip position
and Re $A$ two zeros, one at small values of the momentum transfer (in accordance 
with a theorem by A. Martin for even amplitudes \cite{martinzero}) and
another one around $q^2$ = 2 GeV$^2$.
We also note that the large uncertainty regions associated with Re $A$, as compared
with those from Im $A$, are consequence of the derivative term in parametrization
(13) and the variances/covariances involved. 
Also, from the plots we see that the uncertainty regions from Re $A$ 
are larger at the lower energies, which is explained by our fit procedure
that started at 52.8 GeV, and also the derivative term referred to above.

At last it should be noted that some characteristics of the fit results here presented 
and also of those in our previous empirical analysis \cite{fms11,am} are consequences
of the data ensemble considered, the choice of the analytical parametrization as sum of 
exponential in $q^2$ and the use of statistical errors only and not systematic ones. 
In this respect three comments are in order as follows.

(1) All our analyses have been based on the addition of the data at 27.4 GeV to the 
five ISR sets (23.5 - 62.5 GeV). The empirical justification for this procedure has been 
discussed in detail by \'Avila and Menon in \cite{am}, Sect. 3.3.2. Specifically, 
by testing different cutoffs in the momentum transfer, it has been verified that 
for $q^2 \geq 3.5$ GeV$^2$ all the above differential cross section data follows 
a power law in $q^2$ with power $\lambda = - 7.85 \pm 0.04$ \cite{am}, indicating therefore 
an energy-independent behavior of the differential cross section in this kinematic region. 
On the other hand, in the QCD context, 
it is expected that this region of large momentum transfer in elastic 
scattering may be accessed by perturbative techniques. In this respect the above result 
favors the triple-gluon exchange picture by Donnachie and Landshoff \cite{dl1,dl2}, which 
predicts $\lambda = - 8$, but not the constituent 
interchange model (CIM) by Lepage and Brodsky \cite{lb}, since in this case it is predicted $\lambda = - 10$.
This approach also predicts a scaling of the form 
\begin{eqnarray}
\frac{d\sigma}{dq^2} \sim s^{-10} f(\frac{q^2}{s}),
\nonumber
\end{eqnarray}
which has been verified at lower energies \cite{lb}. There is also indication of agreement with the
data at 19.4 GeV and 27.4 GeV, if a particular power law is assumed for $f(q^2/s)$ and the
systematic uncertainties of $15 \%$ present in both sets are neglected \cite{faissler}.
This small energy dependence in the interface between Fermilab and ISR energies
corroborates the analysis by \'Avila and Menon in the sense to include the data at 27.4 GeV
only at the ISR sets and not at 19.4 GeV (see \cite{am}, Sect. 3.3.2, for more
details).

(2) Another aspect in our analyses concerns the assumption of a sum of exponential in 
$q^2$, for the \textit{whole} range of momentum transfer investigated. This choice has two 
foundations as follows. The first one is strictly empirical since for the differential cross sections
in a logarithm scale all the details of the diffraction peak, the dip-bump structure and the
large $q^2$ region can be well described by compositions of straight lines,
analogous to an exhaustion process (the empirical character of our analysis impose no
limits in the number of free parameters or exponential terms). The second one, to be
treated in the next section, concerns the possibility of \textit{analytical}
translation from the $q^2$-space to the impact parameter space (Fourier transform) and, 
more importantly, the \textit{analytical} evaluation of the uncertainties in some extracted 
quantities, through error propagation from the fit parameters. On the other hand,
as referred to above, the justification for the inclusion of the data at 27.4 GeV
to the ISR sets was based on a power law fit analysis at the large momentum transfer
region and not exponential forms. In this respect, it should be noted that as the
momentum transfer increases, a replacement of an exponential decrease of the
differential cross section to a 
power one may be related to a transition from soft to 
hard sectors and in the phenomenological context can be connected with the geometrical 
structure of the Regge trajectories, from linear to non-linear one, respectively, as 
discussed by Fiore \textit{et al}. \cite{fiore} (Sect. 6). Therefore, it might also be important to 
investigate the possibility to combine exponential and power forms in an unique analytical 
parametrization. However, we shall limit our analyses to exponential forms for the reasons 
referred to above, namely they represent a suitable empirical ansatz, ``unbiased by any 
theoretical prejudice", as suggested by Fiore \textit{et al}. \cite{fiore} and allows analytical 
translation to the impact parameter space.

(3) The last aspect concerns the dataset considered \cite{fms11,am} and the use of statistical 
errors only and not systematic ones. In this respect it should be noted that another data sets exist, 
as that compiled and analyzed by Cudell, Lengyel and Martynov \cite{clm} (where the systematic 
errors are taken into account) and available in \cite{cudell}. We think it would 
also be interesting to develop new data reductions with this data set and our proposed 
parametrization; a comparative analysis on all the results could be useful.

\section{Extracted Overlap and Eikonal Functions}

In this section we consider some implications of the fit results in what concerns
properties of the extracted overlap functions and the imaginary 
part of the eikonal (opacity) in the momentum transfer space. We first recall the main formulas
associated with the impact parameter and eikonal representations (Sect. IV.A)
and then discuss in some detail the extracted inelastic overlap function
(Sect. IV.B) and the opacity function in the momentum transfer space (Sect. IV.C).

\subsection{Impact Parameter and Eikonal Representations}

This subject is treated in detail in \cite{pred}. In what follows we only recall
the main formulas of interest here.

The representation of the elastic scattering
amplitude in the impact parameter space is named profile function and in case of azimuthal symmetry they 
are connected by
\begin{eqnarray}
A(s,q) = \mathrm{i} \int_{0}^{\infty} b\,db\, J_{0}(qb)\, \Gamma(s, b),
\end{eqnarray}
where $b$ is the impact parameter, $\Gamma(s, b)$ the profile function and 
$J_0$ is the zero-order Bessel function.
In the eikonal representation the profile is given by
\begin{eqnarray}
\Gamma(s, b) = 1 - e^{\mathrm{i}\,\chi (s,b)},
\end{eqnarray}
where $\chi (s,b)$ is the eikonal function.

The unitarity principle in the impact parameter space is usually expressed in terms
of the total, elastic and inelastic overlap functions,

\begin{eqnarray}
G_{\mathrm{tot}}(s, b) = G_{\mathrm{el}}(s, b) + G_{\mathrm{in}}(s,b),
\end{eqnarray}
which, in terms of the profile function reads 

\begin{eqnarray}
2\rm{Re} \Gamma(s,b) = | \Gamma(s,b) |^2 + G_{\mathrm{in}}(s, b),
\end{eqnarray}
where $G_{\mathrm{in}}(s, b)$ is the inelastic overlap function and it follows that
the integrated inelastic cross section is given by
\begin{eqnarray}
\sigma_{\mathrm{in}}(s) = 2\pi \int_0^{\infty} b\,db\, G_{\mathrm{in}}(b,s).
\end{eqnarray}

In the eikonal representation, from Eqs. (19) and (21),
\begin{eqnarray}
G_{\mathrm{in}}(s, b) = 1 - e^{- 2\,\mathrm{Im}\ \chi (s,b)}.
\end{eqnarray}
Since unitarity  implies $\mathrm{Im}\ \chi (s,b) \geq 0$ we have
$G_{\mathrm{in}}(s, b) \leq $1, so that from (22) $G_{\mathrm{in}}$ can be interpreted as the probability
of an inelastic event to take place at given $b$ and $s$; in the black disk limit 
$G_{\mathrm{in}} \rightarrow 1$.

From the above result, the imaginary part of the eikonal is associated with
the absorption in the scattering process and for that reason is
named \textit{opacity function}, which we shall denote 
\begin{eqnarray}
\Omega(s,b) \equiv \mathrm{Im} \chi(s,b).
\end{eqnarray}
In the momentum transfer space it is given by the symmetrical Fourier-Bessel transform,
with notation
\begin{eqnarray}
\tilde{\Omega}(s,q) = \int_{0}^{\infty} b\,db\, J_{0}(qb)\, \Omega(s,b).
\end{eqnarray}
Eikonal models can be classified or distinguished according to different
choices for the opacity function in the momentum transfer space, which 
is expected to be connected 
with form factors and elementary amplitudes.

Due to the importance of the physical interpretations associated with the \textit{inelastic overlap
function} and the \textit{opacity function in the momentum transfer space}, we shall display and
discuss these extracted quantities from the fit results. However, before that, we stress again that
the relatively small interval investigated ($\approx$ 19 - 60 GeV) and the unconstrained fit
procedure, turn out difficult to infer energy dependencies on these quantities since, in general,
the error propagation leads to an overlapping of the uncertainty regions. We understand
that this does not constitute a serious drawback in our analyses because
in the phenomenological context what is theoretically unknown,
in general,
concerns the impact parameter and momentum transfer distributions and not the
dependencies on the energy \cite{pred,land,fiore}.
Therefore we shall limit
the results to typical cases of interest at fixed energy, or those for which there is no
overlapping of the uncertainties.

\subsection{Inelastic Overlap Function}

With parametrization (13), the profile function can be {\it analytically evaluated} through
the inverse (symmetrical) transform of Eq. (18),
\begin{eqnarray}
\Gamma(s, b) = - \mathrm{i} \int_{0}^{\infty} q\,dq\, J_{0}(qb)\, A(s,q),
\end{eqnarray}
as well as the overlap functions in Eqs. (20) and (21),
together with the associated uncertainties through error propagation.
For each $b$-interval considered for the extracted quantities, we have generated $10^3$ empirical 
points in the form
of a central value and the corresponding uncertainty, namely $G \pm \Delta G$.

The results for the extracted overlap functions in all the energies investigated are very similar 
and are illustrated in Fig. 3 in the case
of $\sqrt{s}$ = 52.8 GeV (largest set of experimental data). The figure, in linear scale, shows the 
uncertainty regions for the extracted total, elastic and inelastic
overlap functions. We note that at $b=0$, $G_{\mathrm{in}}$
reaches $\approx$ 92\% of the black disc limit, as expected \cite{ag,mit}.

\begin{figure}[pb]
\centering
\vspace*{1.0cm}
\epsfig{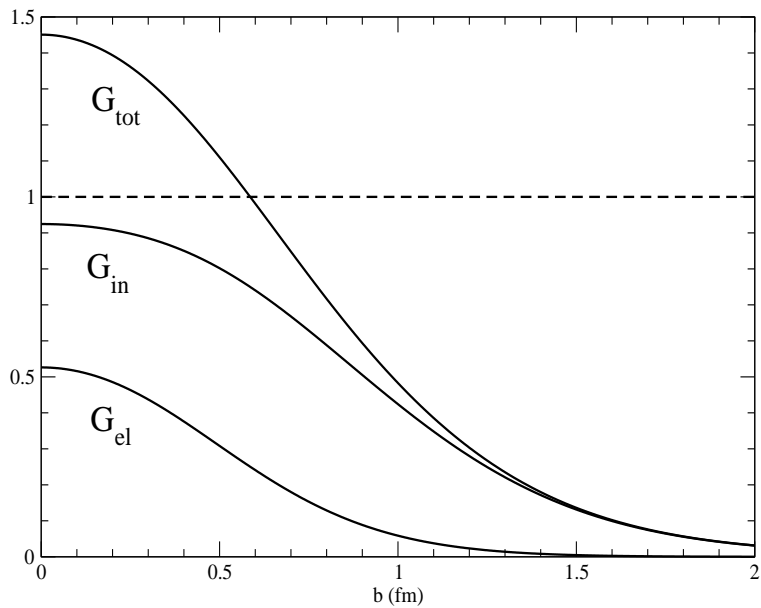}
\vspace*{8pt}
\caption{Uncertainty regions of the extracted overlap functions for $pp$ scattering
at 52.8 GeV and the black disc limit (dashed line).
\label{f3}}
\end{figure}

We now focus on the extracted inelastic overlap function.
In order to develop some test on the reliability of the extracted results,
we have evaluated the integrated inelastic cross section through Eq. (22),
a quantity that was not used as experimental input in the data reductions.
The results for all the sets analyzed are displayed in Table II, showing
plenty agreement with the corresponding experimental data. 
The data at 19.4 GeV has been evaluated from 
$\sigma_{tot}$ = 38.98 $\pm$ 0.04 mb obtained by Carroll {\it et al}. \cite{carroll} and
$\sigma_{el}$ = 6.78 $\pm$ 0.13 mb by Schiz {\it et al}. \cite{schiz}. Those
at the ISR region come from the analysis by Amaldi and Schubert \cite{as}.

\begin{table}[!]
\caption{Integrated inelastic cross section evaluated through Eq. (22) with the
extracted inelastic overlap functions and the experimental data \cite{as,carroll,schiz}.}
{\begin{tabular}{@{}ccc@{}}
\toprule
$\sqrt{s}$ &  $\sigma_{\mathrm{in}}^{fit}$  &  $\sigma_{\mathrm{in}}^{exp}$ \\
   (GeV) &    (mb)     & (mb) \\
\colrule
19.4 &\ 32.03$\pm$0.04 &\ 32.11$\pm$0.14 \\
23.5 &\ 32.21$\pm$0.16 &\ 32.21$\pm$0.14 \\
30.7 &\ 32.99$\pm$0.13 &\ 32.98$\pm$0.14 \\
44.7 &\ 34.62$\pm$0.11 &\ 34.62$\pm$0.14 \\
52.8 &\ 35.23$\pm$0.13 &\ 35.22$\pm$0.16 \\
62.5 &\ 35.64$\pm$0.16 &\ 35.66$\pm$0.21 \\
\botrule
\end{tabular}
\label{ta2}}
\end{table}

Next, let us discuss the structure of $G_{\mathrm{in}}$ in the impact parameter
space.
The result at $\sqrt{s}$ = 52.8 GeV, in the form of points with errors and in a logarithm scale,
is displayed in Figure 4. First we note that in the region 0 - 2 fm
the empirical points follow a Gaussian dependence with center at the
origin, but above this region the decrease is slower and deviate from
the Gaussian in the form of a tail. This effect is not new since it
has already been observed by Amaldi and Schubert \cite{as} and also
Henyey, Tuan and Kane \cite{htk} in the seventies. In terms of the
differential cross section data it is connected with the change of
the slope at $\approx$ 0.13 GeV$^2$, referred to in Sect. III.B.

In order to go further, looking for some quantitative information on this tail effect,
we have tested different analytical parametrization for the empirical
points in Fig. 4 and in this case we have also considered all the energies investigated.
Making use of the CERN-Minuit code and
using as test functions exponential, Gaussian and combinations
of these functions,
the best global result has been obtained with a composition of three Gaussian,
which we shall denote
\begin{eqnarray}
G_{in}(b) = \sum_{i=1}^3 A_i e^{-B_i(b-C_i)^2} = \sum_{i=1}^3 G_i(b),
\end{eqnarray}
where $A_i$, $B_i$ and $C_i$, $i$ = 1, 2, 3 are free fit parameters. The result of the fit,
through the CERN-Minuit code,
for $pp$ at 52.8 GeV is shown in Fig. 4, together with the corresponding components
$G_i(b)$, $i$ = 1, 2, 3; the values of the parameters are displayed in Table III.
In this case, since the ensemble corresponds to extracted empirical points
with errors and not to experimental data we did not evaluate the uncertainties
associated with these components and make no reference to the goodness
of fit (reduced $\chi^2$).

\begin{figure}[pb]
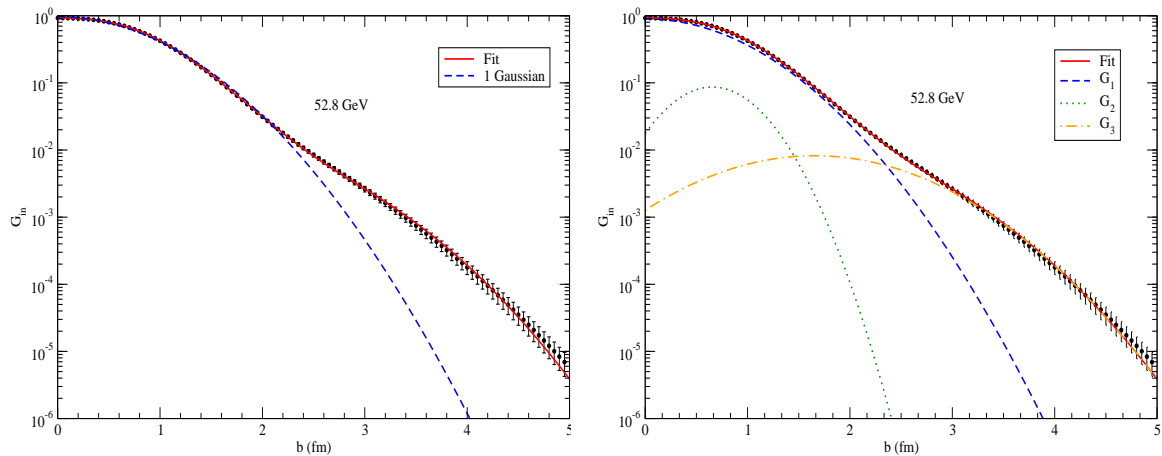

\centering
\vspace*{1.0cm}
\epsfig{file=f4a.eps,width=7.5cm,height=6cm}\hspace{0.2cm}
\epsfig{file=f4b.eps,width=7.5cm,height=6cm}
\vspace*{8pt}
\caption{Extracted points with uncertainties for the inelastic overlap function at 52.8 GeV.
Left: fit with one Gaussian centered at the origin (dashed) and parametrization (27) (solid).
Right: result of the fit and the three Gaussian components in Eq. (27).
\label{f4}}
\end{figure}

\begin{table}[!]
\caption{Fit results for the Gaussian components of the inelastic overlap function,
Eq. (27) at 52.8 GeV.}
{\begin{tabular}{@{}cccc@{}}
\toprule
$i$\ &\  $A_i$  &\  $B_i$ &\ $C_i$ \\
    &         &\ (fm$^{-2}$)&\ (fm) \\
\colrule
$1$\ &\ 0.9005  $\pm$ 0.0067  &\ 0.908 $\pm$ 0.020  &\ 0.0   $\pm$ 0.0  \\
$2$\ &\ 0.0866  $\pm$ 0.0048  &\ 3.71  $\pm$ 0.32   &\ 0.655 $\pm$ 0.015 \\
$3$\ &\ 0.00821 $\pm$ 0.00020 &\ 0.684 $\pm$ 0.024  &\ 1.653 $\pm$ 0.034  \\
\botrule
\end{tabular}
\label{ta3}}
\end{table}

\begin{figure}[pb]
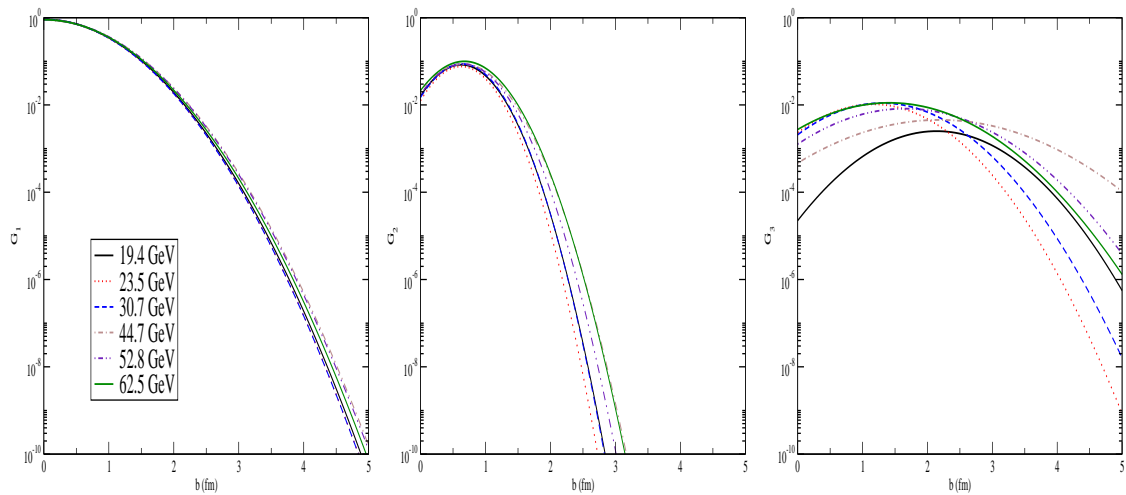

\centering
\vspace*{1.0cm}
\epsfig{file=f5a.eps,width=4.8cm,height=6.5cm}\hspace{0.1cm}
\epsfig{file=f5b.eps,width=4.8cm,height=6.5cm}\hspace{0.1cm}
\epsfig{file=f5c.eps,width=4.8cm,height=6.5cm}
\vspace*{8pt}
\caption{Gaussian components of the inelastic overlap function, Eq. (27), for each set analyzed.
Solid lines correspond to the extrem energies (19.4 and 62.5 GeV) and the other symbols to
the intermediate energies.
\label{f5}}
\end{figure}

The results for each Gaussian component, $G_i$, $i$ = 1, 2, 3, in each energy investigated are displayed in Fig. 5. Although, once more, a dependence on the
energy cannot be extracted in this interval, we can identify the following
remarkable features: (1) each component is concentrated in specific regions of the
impact parameter space, with centers at $b=0$ ($G_1$), around
$b=0.6$ fm ($G_2$) and roughly around
$b=2.0$ fm ($G_3$); (2) $G_1$ dominates the central and intermediate regions and 
$G_3$ describes the tail effect above $b \approx$ 2.2 fm; (3) the effect of different energies is more evident
in $G_3$ then in the others components.

In order to get some quantitative information on the relative contribution of each Gaussian component 
in the integrated
cross section, we have evaluated the following partial quantities:

\begin{eqnarray}
\sigma_{\mathrm{in}}^{G_i}(s) = 2\pi \int_0^{\infty} bdb G_i(b,s),
\qquad
i= 1, 2, 3.
\end{eqnarray}
The results with the corresponding sum of the partial contributions in each energy
are shown in Table IV. Certainly this sum of the areas under the curves 
cannot be identified with
$\sigma_{\mathrm{in}}$, but allows to infer that, roughly, the relative contributions
read 85 \% from $G_1$,  10 \% from  $G_2$ and 5 \% from $G_3$. Moreover, from Table IV, in this energy interval
the relative errors associated with the average of each contribution, namely 
$\Delta<\sigma_{\mathrm{in}}^{G_i}>/<\sigma_{\mathrm{in}}^{G_i}>$, correspond to 3\%, 19\% and 35\%
for $i=1, 2, 3$ respectively. These quantitative results can be interpreted as evidence
that the peripheral region is more sensitive to variations of the energy than the intermediate
region and this one yet more sensitive than the central region. That seems a reasonable result,
since we are treating the inelasticity associated with the elastic channel
through unitarity.
Also, from Table IV we can see that except for the result at
44.7 GeV, the contribution from $G_3$ increases with the energy.

\begin{table}[!]
\caption{Contribution from each Gaussian component to the integrated inelastic cross section
and the sum of the contributions.}
{\begin{tabular}{@{}ccccc@{}}
\toprule
$\sqrt{s}$ &\ \ $\sigma_{\mathrm{in}}^{G_{1}}$ &\ \ $\sigma_{\mathrm{in}}^{G_{2}}$ &\ \ $\sigma_{\mathrm{in}}^{G_{3}}$ &\ $\sum_{i=1}^{3}\sigma_{\mathrm{in}}^{G_{i}}$ \\
    (GeV)  &\ \  (mb)      &\ \ (mb) &\ \ (mb) &\ (mb) \\
\colrule
19.4 & 29.67 & 2.87 & 0.59 & 33.13 \\
23.5 & 29.02 & 2.48 & 1.29 & 32.79 \\
30.7 & 29.23 & 3.01 & 1.63 & 33.87 \\
44.7 & 31.02 & 3.52 & 1.60 & 36.14 \\
52.8 & 31.15 & 3.31 & 1.84 & 36.30 \\
62.5 & 30.26 & 4.26 & 2.12 & 36.49 \\
\botrule
\end{tabular}
\label{ta4}}
\end{table}

In conclusion, these results indicate three Gaussian components for the inelastic overlap function
associated with central, intermediate and peripheral regions. Although an explicitly dependence on the energy
cannot be obtained the results suggest that the peripheral region is more sensitive
to the energy evolution then the central region. We understand that the correct evaluation 
of $\sigma_{\mathrm{in}}$ (Table II)
and the reproduction of the tail effect in $G_{\mathrm{in}}$ give support to our representation
and fit procedure.

\subsection{Opacity Function in the Momentum Transfer Space}

With the extracted profile function we determine the opacity function in
the impact parameter space {\it analytically}, through Eqs. (19) and (24),
\begin{eqnarray}
\Omega(s,b) =
 \ln\left\lbrace \dfrac{1}{\sqrt{[1-\mathrm{Re}\,\Gamma(s,b)]^{2} + [\mathrm{Im}\,\Gamma(s,b)]^{2}}}\right\rbrace,
\end{eqnarray}
together with the corresponding uncertainties by error propagation from the
fit parameters.

Now, since $\Gamma(s,b)$, from $A(s,q)$ in (13), is given by a sum of Gaussian in $b$, the translation
to $q^2$-space, through Eq. (25), cannot be analytically performed and therefore the error propagation neither.
For that reason we have used the semi-analytical approach discussed
by \'Avila and Menon \cite{am} which is valid under certain condition
(as explained in what follows) and that we shall name {\it conditioned expansion method}.

From the fit results and in all the energies analyzed, we have found  that
\begin{eqnarray}
\left[ \frac{\mathrm{Im}\,\Gamma(s,b)}{1-\mathrm{Re}\,\Gamma(s,b)}\right]^{2}
\equiv r(s,b)  \ll 1,
\end{eqnarray}
including the uncertainty regions.
This is illustrated in Fig. 6 in the typical case of
$\sqrt{s}$ = 52.8 GeV. Therefore, Eq. (29) for the opacity can be approximated by
\begin{eqnarray}
\Omega(s,b) \approx  \ln \left\{ \frac{1}{1-\mathrm{Re}\,\Gamma}\right\}.
\end{eqnarray}

\begin{figure}[pb]
\centering
\vspace*{1.0cm}
\epsfig{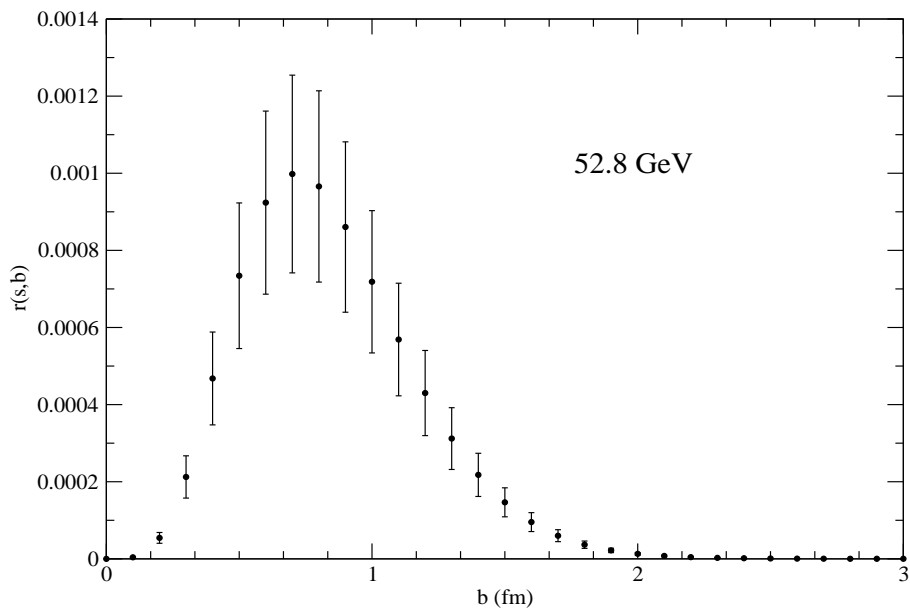}
\vspace*{8pt}
\caption{Empirical points at 52.8 GeV for the ratio $r(s, b)$ in Eq. (30) and uncertainties from
error propagation.
\label{f6}}
\end{figure}

In order to avoid loss of information, we look for what can be directly obtained
from the fit. In this case, by adding and subtracting Re $\Gamma$, the above
equation can be expressed by
\begin{eqnarray}
\Omega(s,b) = \mathrm{Re}\,\Gamma(s,b) + R(s,b),
\end{eqnarray}
where the ``residual"\ function is given by
\begin{eqnarray}
R(s, b) = \left[ \ln \left\{ \dfrac{1}{1-\mathrm{Re}\,\Gamma}\right\} - \mathrm{Re}\,\Gamma(s,b)\right] .
\end{eqnarray}
Now, the Fourier-Bessel transform of (32) reads
\begin{eqnarray}
\tilde{\Omega}(s,q) = \mathrm{Im}A(s,q) + \tilde{R}(s,q),
\end{eqnarray}
so that the first term is directly obtained from the fit and the residual
function can be determined through a {\it semi-analytical method} \cite{am}. Specifically,
with (33) we generate a set of points with errors (from propagation),
$R_i(s,b) \pm \Delta R_i(s,b)$,
$i = 1, 2, ..N$ 
and parametrize this set by Gaussian forms,
\begin{eqnarray}
R(s,b) = \sum_{j=1}^{m} A_{j} e^{-B_{j}b^{2}},
\end{eqnarray}
which has analytical Fourier-Bessel transform and allows error propagation.
Then, substitution in
(34), together with Im$A$ from the fit, leads to $\tilde{\Omega}(s,q) \pm \Delta\tilde{\Omega}(s,q)$.

Following the procedure by \'Avila and Menon \cite{am} we display the results in terms
of $q^8$ multiplied by $\tilde{\Omega}(q)$ together with the uncertainty regions, as shown in Fig. 7
for all sets investigated. A remarkble feature in these results is the evidence of a zero
(change of sign) in the opacity at finite values of the momentum transfer (around 6.5 GeV$^2$),
since even the uncertainties in $q^8\,\tilde{\Omega}(s,q)$ lie below the zero above the crossing
point. Beyond this change of sign, the opacity goes smoothly to zero through
negative values.

\begin{figure}[pb]
\centering
\epsfig{file=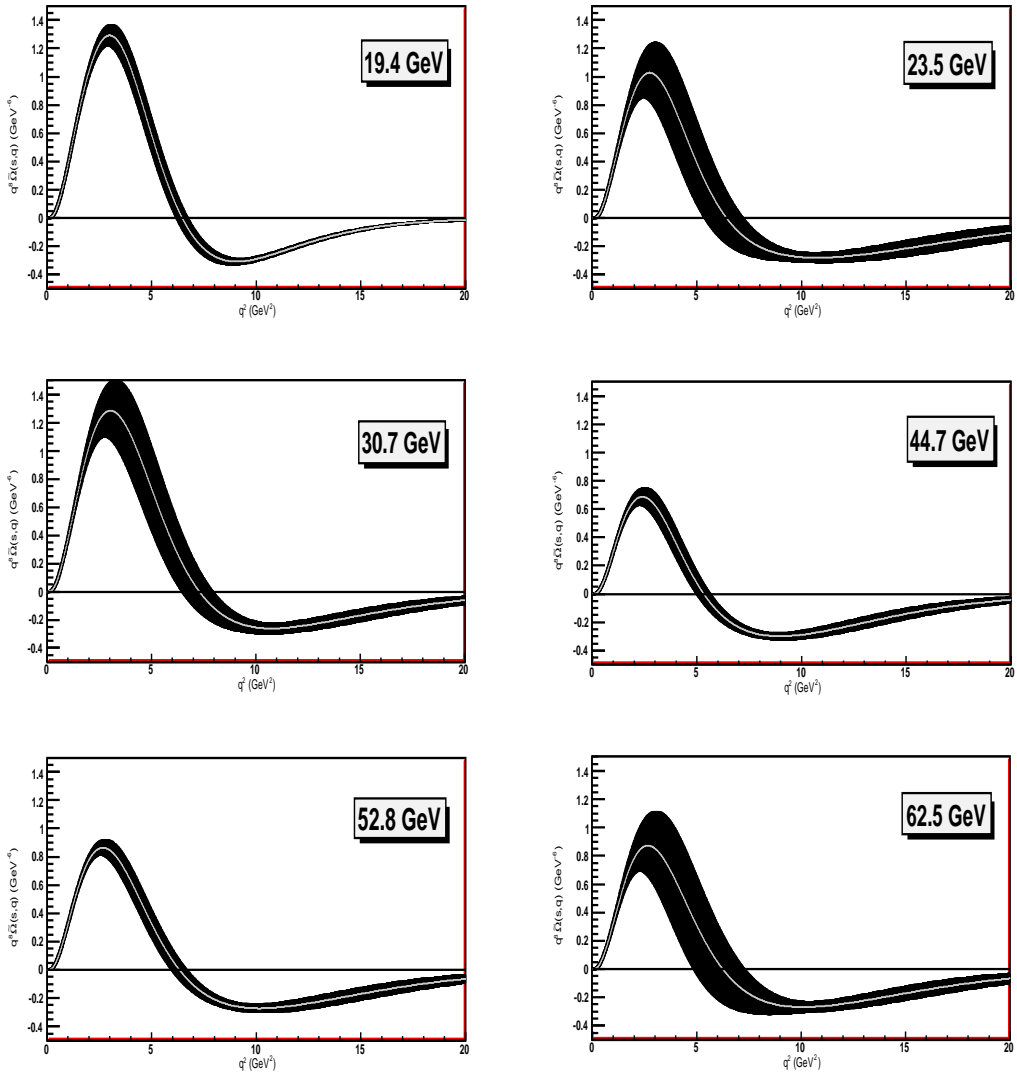,width=14.5cm,height=15cm}
\vspace*{8pt}
\caption{Extracted opacity function in the momentum transfer space multiplied
by $q^8$ and uncertainty regions from error propagation.
\label{f7}}
\end{figure}

From the plots in Fig. 7 the position of the zero can be determined and the error involved
estimated from the uncertainties in both sides of the zero position. The results for
these zero positions in terms of the energy are displayed in Fig. 8. We note that
in the previous analysis by \'Avila and Menon with the constrained parametrization \cite{am} there
has been no evidence for a zero at 19.4 GeV. That is a consequence of the new compilation
and normalization for this data set used here and described in detail in \cite{fms11}.

\begin{figure}[pb]
\centering
\vspace*{1.0cm}
\epsfig{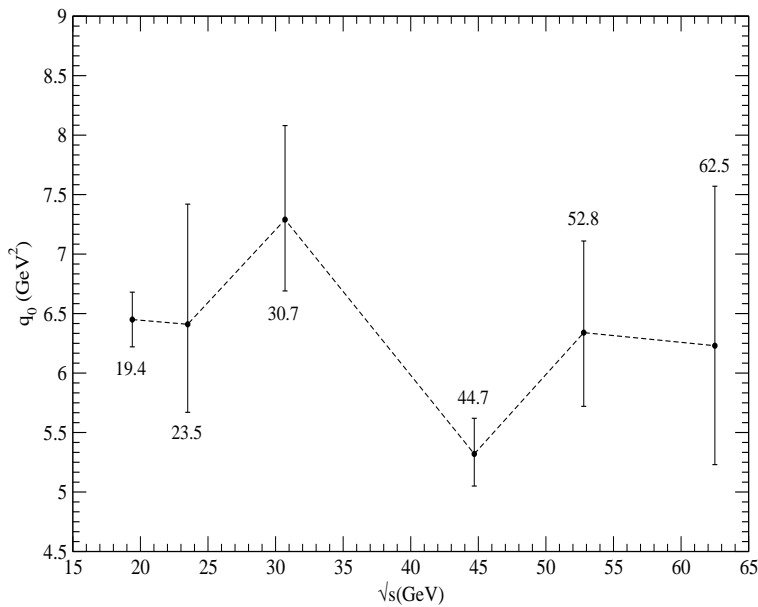}
\vspace*{8pt}
\caption{Position of the zero in the opacity function extracted from the plots
in Fig. 7 (see text) in terms of the energy. The dashed lines are drawn only to guide the eyes.
\label{f8}}
\end{figure}

In order to obtain an analytical parametrization for  the empirical points
displayed in Fig. 7,
we consider a general structure inspired in eikonal models with zero in the
opacity function \cite{elam1,elam2}
and based on the Glauber multiple diffraction
formalism \cite{glauber1,glauber2,glauber3}.
For fixed $s$ we express:

\begin{eqnarray}\label{eq:omega}
\tilde{\Omega}(q) = C G^2(q) f(q),
\end{eqnarray}
with
\begin{eqnarray}
G(q)=\dfrac{1}{(1+q^{2}/\alpha^{2})}\,\dfrac{1}{(1+q^{2}/\beta^{2})}
\end{eqnarray}
and
\begin{eqnarray}\label{eq:f}
f(q) = \frac{1 - q^2/q_0^2}{1 + (q^2/q_0^2)^n},
\end{eqnarray}
where $C$, $\alpha$, $\beta$, $q_0$ are free fit parameters and $n$ a fixed integer.
The change of sign (zero) in $\tilde{\Omega}(q)$ is generated at $q^2 = q_0^2$.

These are general and useful formulas, since for different values of the integer $n$
in $f(q)$ they reproduce different model assumptions or empirical parametrization
for the corresponding opacity function. In fact, for $n=1$ we have a structure used in the Bourrely, 
Soffer and Wu model \cite{bsw1,bsw2,bsw3,bsw4}, which we shall denote $\tilde{\Omega}_{BSW}$. The case
$n=2$ corresponds to a modification in the BSW function, which appears in multiple 
diffraction \cite{menon1,menon2,menon3} and hybrid \cite{covolan} models and we shall denote
$\tilde{\Omega}_{mBSW}$ (standing for {\it modified}-BSW). At last, in the analysis
by \'Avila and Menon the value $n=4$ has been inferred on {\it empirical} basis and for that
reason we shall denote $\tilde{\Omega}_{empir}$.

As before, in what follows we shall focus on the typical results obtained at $\sqrt{s}$ = 52.8 GeV.
The point is to fit the empirical points with errors for the opacity function,
as displayed in Fig. 7, through parametrization (36-38), in the cases
$n=1$ ($\tilde{\Omega}_{BSW}$), $n=2$ ($\tilde{\Omega}_{mBSW}$) and
$n=4$ ($\tilde{\Omega}_{empir}$). In this procedure we have fixed the position of the zero
to the empirical value at this energy, namely $q_0^2$ = 6.34 GeV$^2$ and let free the
parameters $C$, $\alpha^2$ and $\beta^2$.
The results of the fits are displayed in Table V
and Fig. 9, for $q^8\tilde{\Omega}$ and also the modulus $|\tilde{\Omega}|$.

We conclude with this analysis that only $\tilde{\Omega}_{empir}$ can reproduce the empirical
points and in this case, from Table 5, $\beta^2 \approx 1 + \alpha^2$ (a form factor discussed 
by G. Shaw, in the beginning of the seventies \cite{shaw1,shaw2}). These results are
in plenty agreement with those obtained by \'Avila and Menon through
the constrained parametrization \cite{am} and favor models with a zero in the eikonal
(see \cite{am} for more details).

\begin{figure}[pb]
\centering
\vspace*{1.0cm}
\epsfig{file=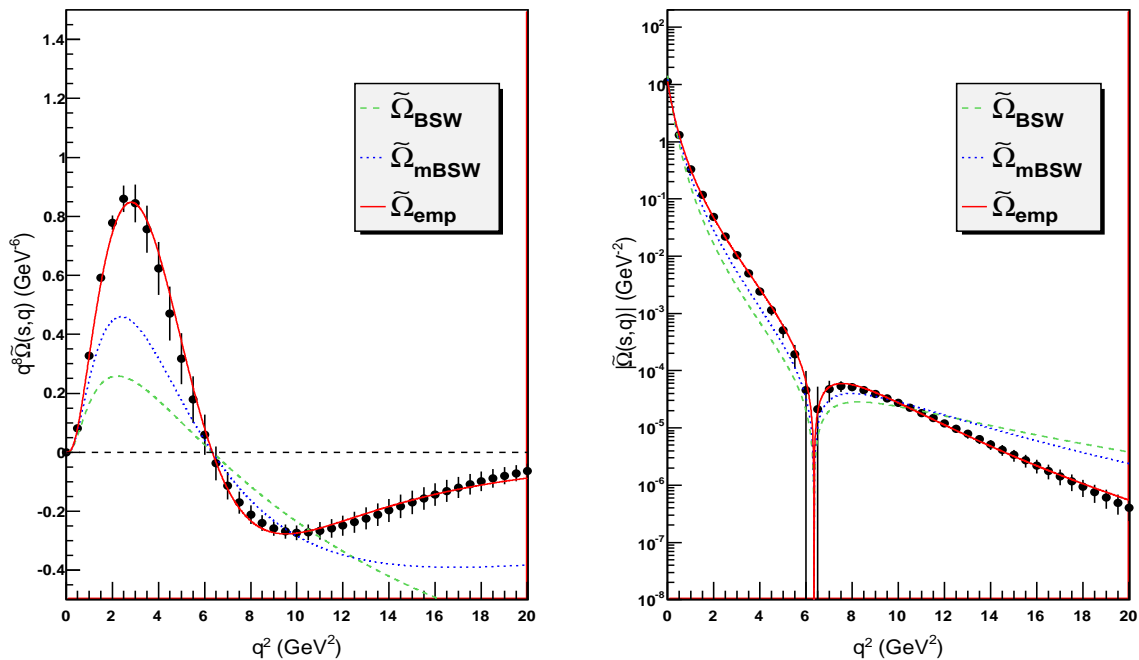,width=16.0cm,height=10cm}
\vspace*{8pt}
\caption{Extracted opacity function at $\sqrt{s}$ = 52.8 GeV and parametrization 
through Eqs. (36-38) for different values of $n$ in Eq. (38). Numerical and statistical 
information are shown in Table V.
\label{f9}}
\end{figure}

\begin{table}[!]
\caption{Results of the fits to the empirical points in Fig. 7 through Eqs. (36-38)
at $\sqrt{s}$ = 52.8 GeV.}
{\begin{tabular}{@{}cccc@{}}
\toprule
 parameters          &\ \ $\tilde{\Omega}_{BSW}$ ($n=1$)&\ \ $\tilde{\Omega}_{mBSW}$ ($n=2$)&\ \ $\tilde{\Omega}_{empir}$
($n=4$) \\
   $C$  (GeV$^{-2}$) &\ \ 14.101 $\pm$ 0.024  &\ \ 12.303 $\pm$ 0.022  &\ \ 11.162 $\pm$ 0.031  \\
$\alpha^2$  (GeV$^2$)&\ \ 0.5515 $\pm$ 0.0024 &\ \ 0.6509 $\pm$ 0.0040 &\ \ 0.4622 $\pm$ 0.0041  \\
$\beta^2$  (GeV$^2$) &\ \ 0.5515 $\pm$ 0.0024 &\ \ 0.6509 $\pm$ 0.0040 &\ \ 1.431  $\pm$ 0.018   \\
\hline
     $\chi^2$/DOF    &\ \     136            &\ \     40             &\ \     0.19          \\
\botrule
\end{tabular}
\label{ta5}}
\end{table}

\section{Conclusions and Final Remarks}

The absence of a pure QCD treatment  of the elastic hadron scattering 
and the lack of a widely accepted phenomenological approach to these processes,
have motivated us to investigate the inverse problem, in the context of the 
impact parameter and eikonal representations.
As commented along the paper  two of the main problems with these model-independent analyses 
are: (1) we do not have information on the contributions from the
real and imaginary parts of the amplitude beyond the forward direction;
(2) the fits being non-linear have no unique solution. 
As a consequence, 
different kinds of solutions and contributions from the real and imaginary parts of the amplitude must be tested
and reliable information should not strongly depend on the details of these 
solutions and contributions.

With this strategy in mind, we have already developed two different empirical solutions for
data reductions both statistically consistent and characterized by 
different contributions from the real and imaginary parts \cite{fms11,am}.
Motivated by these differences we have discussed here the applicability of 
a representation for the Martin's formula without the full scaling property,
which predicts also different contributions from the previous analyses.

A detailed review on the original derivation of this formula and critical comments
on its further developments have been presented. The use of the proposed almost
model-independent representation in empirical fits of the differential cross section
data in the interval 19.4 GeV - 62.5 GeV has led to the following main conclusions:

\begin{itemize}

\item  In the interval investigated, the experimental data on the differential cross sections
at $q^2$ = 0 and $q^2 >$ 0.01 GeV$^2$ are quite well described.

\item  The extracted inelastic overlap function cannot be parametrized by one Gaussian distribution
due to the tail effect, associated with the change of the slope in the differential cross section
at small values of the momentum transfer. The empirical results indicate three Gaussian components
whose sensitivity to variations of the energy increases
from the central to the peripheral regions.

\item  The opacity function in the momentum transfer space, extracted through the 
conditioned expansion method,
present a change of sign around 6.5 GeV$^2$ in all sets analyzed. The empirical points are quite well 
described by a parametrization
consisting of a product of two simple constrained poles 
($\beta^2 \approx 1 + \alpha^2$) by the function $f(q)$ with zero and n=4.

\end{itemize}

Despite the promising aspects of the above results we stress that only a detailed comparative investigation, 
based
on different empirical parametrization and solutions and aimed to identify results that are
common to all the analyses, can bring new reliable insights for model developments
and possible connections with QCD. A global critical review along these lines, collecting
all our previous results and those presented here, are being prepared for a forthcoming 
communication.

\section*{Acknowledgments}

This research was supported by FAPESP 
(Contracts Nos. 11/00505-0, 09/50180-0, 07/05953-5).


\begin{thebibliography}{0} 

\bibitem{pred} 
V. Barone and E. Predazzi, {\it High-Energy Particle Diffraction}
(Spring-Verlag, Berlin, 2002).

\bibitem{land}
S. Donnachie, G. Dosch, P.V. Landshoff and O. Natchmann, 
{\it Pomeron Physics and
QCD}, (Cambridge University Press, 2002).

\bibitem{fiore}
R. Fiore, L. L. Jenkovszky, R. Orava, E. Predazzi, A. Prokudin and O. Selyugin, 
{\it Int. J. Mod. Phys. A\/} {\bf 24}, 2551 (2009). 


\bibitem{totem1} 
http://public.web.cern.ch/public/en/lhc/TOTEM-en.html.

\bibitem{totem2} 
J. Kaspar, TOTEM experiment: Elastic and total cross sections,
in {\it Proc. 13th Int. Conf. Elastic and Diffractive Scattering},
eds. M. Deile, D. d'Enterria, A. De Roeck (DESY, Hamburg, 2010), p.~55.

\bibitem{fms11}
D.A. Fagundes, M.J. Menon and G.L.P. Silva, 
{\it Eur. Phys. J. C\/} {\bf 71}, 1637 (2011).

\bibitem{fms10} 
D.A. Fagundes, M.J. Menon and G.L.P. Silva, 
Empirical Results for the Eikonal from Proton-Proton Elastic Scattering.
in {\it Proc. XI Hadron Physics}, eds. M. Nielsen, F.S. Navarra and M.E. Bracco,
AIP Conference Proceedings 1296 (AIP, Melville, 2010), p.~278.


\bibitem{fm10} 
D.A. Fagundes and M.J. Menon, On a Model-independent Representation for the Real Part
of the Elastic Hadron Amplitude,
in {\it Proc. XI Hadron Physics}, eds. M. Nielsen, F.S. Navarra and M.E. Bracco,
AIP Conference Proceedings 1296 (AIP, Melville, 2010), p.~282.

 \bibitem{am}
R.F. \'Avila and M.J. Menon, {\it Eur. Phys. J. C\/} {\bf54}, 555 (2008).

\bibitem{sma}
G.L.P. Silva, M.J. Menon and R.F. \'Avila, {\it Int. J. Mod. Phys. E\/} {\bf 16},
2923 (2007).

\bibitem {acmm} R. F. \'Avila, S. D. Campos, M. J. Menon, and J. Montanha,
{\it Eur. Phys. J. C\/}, {\bf 47}, 171 (2006).

\bibitem{cmm}
P.A.S. Carvalho, A.F. Martini and M.J. Menon, {\it Eur. Phys. J. C\/} {\bf 39},
359 (2005).

\bibitem{cm}
P.A.S. Carvalho and M.J. Menon, {\it Phys. Rev. D\/} {\bf 56},
7321 (1997).

\bibitem{martinf}
A. Martin, {\it Lett. Nuovo Cim.\/} {\bf 7}, 811 (1973).

\bibitem{theo1}
P. Valin, {\it Phys. Rep.\/} {\bf 203}, 233 (1991).

\bibitem{theo2}
J. Fischer, {\it Phys. Rep.\/} {\bf 76}, 157 (1981).

\bibitem{theo3}
S.M. Roy, {\it Phys. Rep.\/} {\bf 5}, 125 (1972).

\bibitem{theo4}
R.J. Eden, {\it Rev. Mod. Phys.\/} {\bf 43}, 15 (1971).


\bibitem{martinmatthiae}
A. Martin and G. Matthiae, Elastic Scattering and Total Cross Section, 
in {\it Proton-Antiproton Collider Physics}, eds. G. Altarelli and L. Di Lella,
Advanced Series on Directions in High Energy Physics, Vol.~4  (World  Scientific,
Singapure, 1989), p.~45.

\bibitem{bc}
M.M Block and R. Cahn, {\it Rev. Mod. Phys.\/} {\bf 57}, 563 (1985) 

\bibitem{fromar1}
M. Froissart, {\it Phys. Rev.\/} {\bf 123}, 1053 (1961).

\bibitem{fromar2}
A. Martin, {\it Phys. Rev.\/} {\bf 129}, 1432 (1963).

\bibitem{fromar3}
A. Martin, {\it Nuovo Cim.\/} {\bf 42}, 930 (1966).

\bibitem{akm}
G. Auberson, T. Kinoshita and A. Martin, {\it Phys. Rev. D\/} {\bf 3}, 3185 (1971).

\bibitem{beg1}
J. Bros, H. Epstein and V. Glaser, {\it Nuovo Cimento\/} {\bf 31}, 1265 (1964).

\bibitem{beg2}
J. Bros, H. Epstein and V. Glaser, {\it Com. Math. Phys.\/} {\bf 1}, 240 (1965).

\bibitem{eden}
R.J. Eden, {\it High Energy Collisions of Elementary Particles}
(Cambridge University Press, Cambridge, 1967), Sect. 7.1.

\bibitem{dd}
J. Dias de Deus, {\it Nucl. Phys. B\/} {\bf 59}, 231 (1973).

\bibitem{gs1}
A.J. Buras and J. Dias de Deus, {\it Nucl. Phys. B\/} {\bf 71}, 481 (1974).

\bibitem{gs2}
J. Dias de Deus, {\it Nuovo Cim. A\/} {\bf 28}, 114 (1975).

\bibitem{gs3}
J. Dias de Deus and P. Kroll, {\it Acta  Physica Polonica B\/} {\bf 9}, 157 (1978).

\bibitem{ddr1}
N.V. Gribov and A.A. Migdal, {\it Sov. J. Nucl. Phys.\/} {\bf 8}, 583 (1969).

\bibitem{ddr2}
J.B. Bronzan, G.L. Kane and U.P. Sukhatme, {\it Phys. Lett. B\/} {\bf 49}, 272 (1974).

\bibitem{ddr3}
R.F. \'Avila and M.J. Menon,  {\it Nucl. Phys. A\/} {\bf 744}, 249 (2004).

\bibitem{bozzo}
UA4 Coll. (M. Bozzo {\it et al}.),  {\it Phys. Lett. B\/} {\bf 147}, 392 (1984).

\bibitem{hv1}
R. Henzi and P. Valin,  {\it Phys. Lett. B\/} {\bf 132}, 443 (1983).

\bibitem{hv2}
R. Henzi and P. Valin, {\it Phys. Lett. B\/} {\bf 149}, 234 (1984).

\bibitem{hv-high}
R. Henzi and P. Valin,  {\it Phys. Lett. B\/} {\bf 160}, 167 (1985).

\bibitem{kl1}
V. Kundr\'at and M. Lokaj\'{\i}\v{c}ek, {\it Phys. Rev. D\/}, {\bf 55}, 3221 (1997).

\bibitem{kl2}
V. Kundr\'at and M. Lokaj\'{\i}\v{c}ek, {\it Phys. Lett. B\/} {\bf 232}, 263 (1989).

\bibitem{kl3}
V. Kundr\'at and M. Lokaj\'{\i}\v{c}ek, {\it Phys. Rev. D\/}, {\bf 31}, 1045 (1985).

\bibitem{kmy-c}
M. Kawasaki, T. Maehara and M. Yonezawa, {\it Phys. Rev. D\/}, {\bf 55}, 3225 (1997).


\bibitem{roy}
S.M. Roy, {\it Phys. Lett. B\/} {\bf 135}, 191 (1984).


\bibitem{bcmp}
J. Bellandi, R.J.M. Covolan, M.J. Menon and B.M. Pimentel, {\it Hadronic J.} {\bf 11},
17 (1988).

\bibitem{malecki}
A. Ma{\l}ecki, {\it Phys. Lett. B\/} {\bf 149}, 221 (1989).


\bibitem{menon1} 
M.J. Menon and B.M. Pimentel, {\it Hadronic J.} {\bf 13},
325 (1990).

\bibitem{cy1}
T.T. Chou and C.N. Yang, {\it Phys. Lett. B\/} {\bf 244}, 113 (1990).

\bibitem{cy2}
T.T. Chou and C.N. Yang, {\it Phys. Lett. B\/} {\bf 128}, 457 (1983).

\bibitem{menon2}
M.J. Menon, {\it Nucl. Phys. B. (Proc. Suppl.)}, {\bf 25}, 94 (1992).

\bibitem{menon3}
M.J. Menon, {\it Can. J. Phys.}, {\bf 74}, 594 (1996).

\bibitem{kmy1}
M. Kawasaki, T. Maehara and M. Yonezawa, {\it Phys. Rev. D\/}, {\bf 47}, R3 (1993).

\bibitem{kmy2}
M. Kawasaki, T. Maehara and M. Yonezawa, {\it Phys. Rev. D\/}, {\bf 48}, 3098 (1993).

\bibitem{ddp}
J. Dias de Deus and A.B. P\'adua, {\it Phys. Lett. B\/} {\bf 317}, 428 (1993).

\bibitem{as}
U. Amaldi and K.R. Schubert, {\it Nucl. Phys. B\/} {\bf 166}, 301 (1980).

\bibitem{nagy}
E. Nagy {\it et al}., {\it Nucl. Phys.  B\/}, {\bf 150}, 221 (1979).


\bibitem{faissler}
W. Faissler {\it et al}., {\it Phys. Rev. D\/}, {\bf 23}, 33 (1981).

\bibitem{dl1}
A. Donnachie and P.V. Landshoff, {\it Z. Phys.  C\/}, {\bf 2}, 55 (1979).

\bibitem{dl2}
A. Donnachie and P.V. Landshoff, {\it Phys. Lett. B\/} {\bf 387}, 637 (1996).

\bibitem{minuit}
F. James,
MINUIT - Function Minimization and Error Analysis (CERN, 2002)

\bibitem{bevington}
P.R. Bevington and D.K. Robinson, {\it Data Reduction and Error Analysis
for the Physical Sciences}
(McGraw-Hill, Boston, Massachusetts, 1992).

\bibitem{martinzero}
A. Martin, {\it Phys. Lett. B\/} {\bf 404}, 137 (1997).

\bibitem{lb}
G.P. Lepage and S.J. Brodsky, {\it Phys. Rev. D\/}, {\bf22}, 2157 (1980).

\bibitem{clm}
J.R. Cudell, A. Lengyel and E. Martynov, Phys. Rev. D \textbf{73}, 034008 (2006)

\bibitem{cudell}
http://nuclth02.phys.ulg.ac.be/cudell/data


\bibitem{ag}
G. Alberi and G. Goggi, {\it Phys. Rep.} {\bf 74}, 1 (1981).

\bibitem{mit}
H.I. Miettinen, Impact Structure of Diffraction Scattering, in {\it
Proc. 9th Rencontre de Moriand}, ed. J. Tran Than Van (Moriond Conference,1974);
preprint CERN-TH 1864 (1974).

\bibitem{carroll}
A.S. Carroll {\it et al}., {\it Phys. Lett. B\/}, {\bf 80}, 423 (1979).

\bibitem{schiz}
A. Schiz {\it et al}., {\it Phys. Rev. D\/}, {\bf24}, 26 (1981).

\bibitem{htk}
F.S. Henyey, R.H. Tuan and G.L. Kane, {\it Nucl. Phys.  B\/}, {\bf 70}, 445 (1974).

\bibitem{elam1}
M.J. Menon, Phys. Rev. D \textbf{48}, 2007 (1993), Erratum-ibid. D \textbf{51}, 1427 (1995).

\bibitem{elam2}
A.F. Martini, M.J. Menon and D.S. Thober, Phys. Rev. D \textbf{54}, 2385 (1996).

\bibitem{glauber1}
R.J. Glauber, in {\it Lectures in Theoretical 
Physics}, edited by W. E. Brittin {\it et al.} (Interscience, New York, 1959) Vol. I, p.315.

\bibitem{glauber2}
W. Czy\.{z} and L.C. Maximon, Ann. Phys. (N.Y.)
{\bf52}, 59 (1969).

\bibitem{glauber3}
V. Franco and G.K. Varma, Phys. Rev. C {\bf 18}, 349 (1978).


\bibitem{bsw1}
C. Bourrely, J. Soffer and T.T. WU, Eur. Phys. J. C \textbf{28}, 97 (2003).

\bibitem{bsw2}
C. Bourrely, J. Soffer and T.T. WU, Phys. Rev. D {\bf 19}, 3249 (1979).

\bibitem{bsw3}
C. Bourrely, J. Soffer and T.T. WU, Nucl. Phys. \textbf{B247}, 15 (1984). 

\bibitem{bsw4}
C. Bourrely, J. Soffer and T.T. WU, Phys. Rev. Lett. {\bf 54}, 757 (1985).

\bibitem{covolan} R.J.M. Covolan, P. Desgrolard, M. Giffon, L.L. Jenkovszky and
E. Predazzi, Z. Phys. C \textbf{58}, 109 (1993).

\bibitem{shaw1}
G. Shaw, Phys. Lett. B \textbf{39}, 255 (1973).

\bibitem{shaw2}
M. Kac, Nucl. Phys. B \textbf{62}, 402 (1973).

\end{thebibliography}
\end{document}